\documentclass[aps,prb,twocolumn,superscriptaddress,floatfix]{revtex4-2}

\usepackage{multirow, array,amsmath}
\usepackage{verbatim}
\usepackage{graphicx}
\usepackage{color}

\begin{document}
	\title{Phonon dynamics in chromium under pressure: absence of phonon criticality at the approach of the quantum critical point}
	
	\author{P. Rodi\`ere}\email{pierre.rodiere@neel.cnrs.fr}
	\affiliation{Universit\'e Grenoble Alpes, CNRS, Institut N\'eel, 38000 Grenoble, France  }  
	
	\author{J.E. Lorenzo}\email{emilio.lorenzo@neel.cnrs.fr}
\affiliation{Universit\'e Grenoble Alpes, CNRS, Institut N\'eel, 38000 Grenoble, France  }

	\author{Q. N. Meier}
\affiliation{Universit\'e Grenoble Alpes, CNRS, Institut N\'eel, 38000 Grenoble, France  }

	\author{L. Paolasini}
	\affiliation{European Synchrotron Radiation Facility (ESRF), 71 avenue des Martyrs, 38000 Grenoble, France.}
	
	\author{A. Bosak}
	\affiliation{European Synchrotron Radiation Facility (ESRF), 71 avenue des Martyrs, 38000 Grenoble, France.}

\date{\today}
	
	\begin{abstract}
		
 Fermi surface nesting is key to understanding the origin of the itinerant antiferromagnetic spin density wave in chromium. This magnetic order is accompanied by a charge-density wave, and both density waves disappear above the critical pressure $P_c \approx 10$ GPa at low temperatures, defining a quantum critical point. Whether this pressure-induced quantum phase transition is accompanied by a phonon instability remains an open question. Here, we use inelastic x-ray scattering to track the room-temperature acoustic phonon dispersions at pressures up to $P=15.6$ GPa, well above $P_c$. We focus on the Kohn anomalies near the $H$ and $N$ points of the Brillouin zone, with the $H$ point anomaly lying close to the incommensurate spin-density-wave ordering vector. The phonon branches harden smoothly under pressure, while the positions and wave-vector extents of both anomalies remain essentially unchanged across $P_c$, with no additional critical softening. \textit{Ab initio} calculations likewise show that Fermi surface nesting remains robust above $P_c$. These results indicate that the pressure-induced quantum phase transition is not driven by a phonon instability and support a primarily spin-density-wave-driven mechanism in presence of a robust Kohn anomaly.

	\end{abstract}

	\maketitle
	
	\section{Introduction}

	A quantum critical point (QCP) is a phase transition at 0~K as a function of a tuning parameter that can be pressure, magnetic field or doping \cite{Sachdev2011,Continentino2017}. QCPs are of great current interest because of their singular ability to influence the finite temperature properties of materials. The occurrence of this paradigmatic QCP is ubiquitous as it appears in ferromagnetic, antiferromagnetic compounds as well as in superconductor pairing system or as a result of purely electronic interactions. Unconventional superconductivity often emerges in the proximity of a magnetically ordered phase, raising the possibility that  spin fluctuations associated to the proximity of a QCP may mediate the formation of superconducting Cooper pairs.

 Chromium has a simple body centered cubic crystal (bcc) structure and is considered the archetype of an antiferromagnetic spin density wave (SDW). At ambient pressure, the spin density wave develops at $T_N$ = 311~K with a modulation wavevector $\boldsymbol{q}$=(0, 0, 1-$\delta$) and $\delta$=0.037 at $T_N$. The itinerant antiferromagnetism in Cr is the result of many body effects and the value of $\delta$ is associated with the nesting wavevector of nearly octahedral electron and hole Fermi surfaces via Coulomb attraction \cite{Lomer62}. The incommensurability arises from the slightly different sizes of the hole and electron surfaces, as it has been assessed by the change of the modulation wavevector upon doping.

\begin{figure}[htbp]
		\begin{center}
			\includegraphics[width = 0.9\columnwidth]{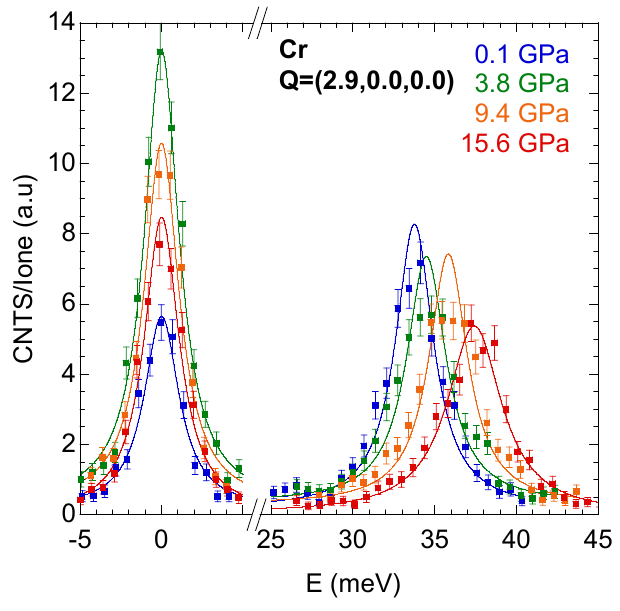}
			\caption{\small Pressure dependence of the longitudinal acoustic phonon mode at 0.9$\Gamma$-H for $\boldsymbol{Q}$=(2.9,0,0). Each point is the number of photon detected on an acquisition time of 2 minutes/point normalized by the incident flux. Solid curves are a fit of the data.} \label{Fig1}
		\end{center}
\end{figure}
	
Elastic neutron scattering experiments have revealed the presence of 3$^{rd}$ order harmonic of the modulation at 3$\delta$ of magnetic origin \cite{Pynn76}. X-ray diffraction experiments sensitive to the charge modulation \cite{Hill95} have shown that 2$^{nd}$ and 4$^{th}$ orders also exist at 2$\delta$ and 4$\delta$. Importantly, the intensities of the $n-$harmonic scale well with that of the first harmonic \cite{Pynn76,Hill95}, $I_{n\delta}(T)\propto I_{\delta}^{n}(T)$. Even-harmonics may result from a strain wave accompanying the onset of the SDW \cite{Young74} which is parented to a charge density wave (CDW) in the language of electronic transitions. In other words, the modulation of the spin amplitude straightforwardly implies a modulation of the effective charge. The understanding of CDW phase transitions largely stems from the work on quasi one dimensional (Q1D) metals where the one dimensionality brings in a metal-insulator phase transition through electron-phonon coupling at a given wavevector \cite{Monceau12}. Marked deviations to a regular phonon dispersion curve at the approach of the CDW phase transition have been reported in many of these Q1D compounds resulting in the Kohn-anomalies or a highly localized softening of a phonon mode of the appropriate symmetry at the position where the incommensurability arises.

In the particular case of Cr, phonon dispersion curves measured well above room temperature have shown the presence of a phonon softening at $\boldsymbol{q}$=(2/3, 2/3, 2/3) upon increasing temperature. This softening is inherent to simple metals having a bcc structure \cite{Trampenau93}. However, it has little to do with a possible electron-phonon interaction leading to the SDW phase transition at $T_N$. Other distinct softenings \cite{Shaw1971,Lamago10} have been observed in the phonon dispersion curve around Brillouin zone boundaries at $\boldsymbol{q}$=(1, 0, 0) (or $H-$point) and $\boldsymbol{q}$=(1/2, 1/2, 0) (or $N-$point). The dip survives in the paramagnetic phase, as well. The only change observed is the overall hardening of the phonon dispersion, which is consistent with extrapolating the results of Refs.\cite{Trampenau93} to lower temperatures. These features have lent support to the occurrence of a phonon-assisted mechanism at the origin of the SDW. Despite the presence of this drop, inelastic X-ray scattering studies on the temperature dependence of the phonon dispersion curves have not reported any change in the position or in the extent of the drop, thus casting doubts on the role of phonons at explaining the appearance of the SDW state \cite{Lamago10}. 

Early studies on the pressure dependence of macroscopic variables in Cr have shown that $T_N \rightarrow 0$ beyond 8~GPa \cite{McWhan1967}, thus supporting the idea of the presence of a QCP. A continuous disappearance of magnetic order near the critical pressure $P_c \approx $10 GPa has been recently confirmed in X-ray diffraction studies of the $P-$dependence of the intensity and q-position of both the spin density wave and charge density wave peaks \cite{Feng07,Jaramillo09}. Close to $P_c$ the quantum fluctuations play a role on the disappearance of this order. The occurrence of a QCP has been confirmed in Cr-V alloys, Cr$_{1-x}$V$_x$ with x$_C$= 0.032, and constitutes a second road to achieve quantum criticality.

X-ray diffraction experiments under pressure of the CDW peaks have shown that the nesting at the Fermi surface improves at the approach of the QCP \cite{Jaramillo09}. One might therefore expect pressure-induced phonon softening to evolve as the QCP is approached. Here, we investigate this pressure dependence in order to reveal potential soft phonon behavior.

\begin{figure*}[tb]
		\begin{center}
			\includegraphics[width = 0.7\columnwidth]{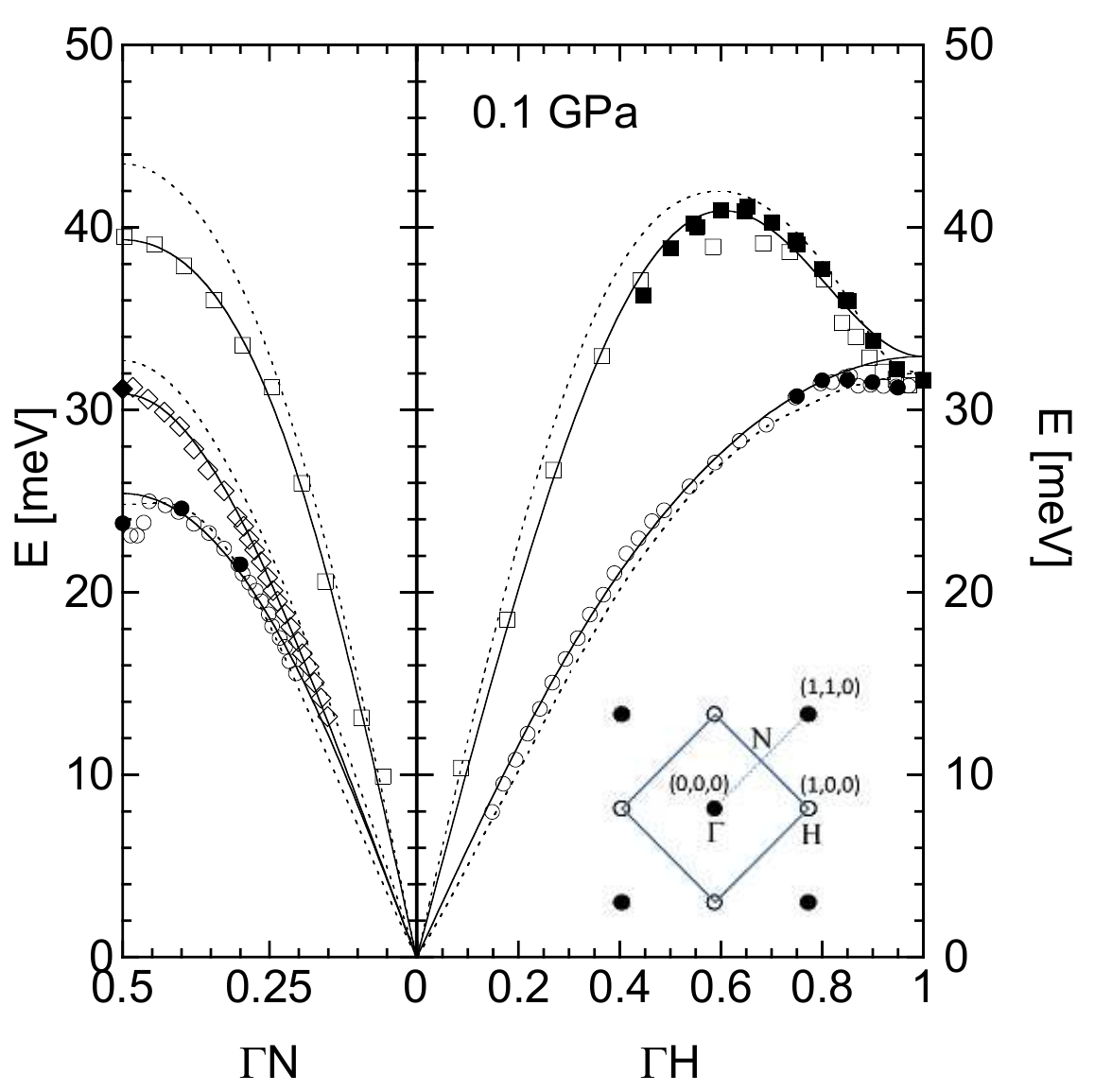}
			\includegraphics[width = 0.72\columnwidth]{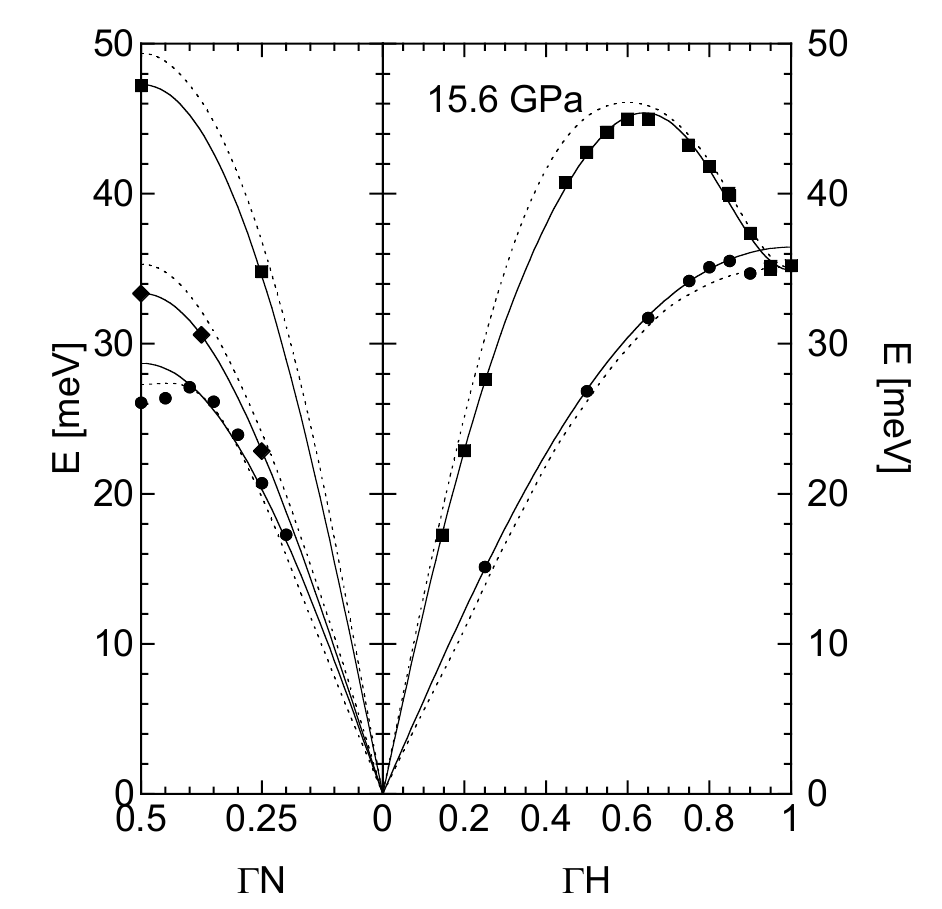}
			\caption{\small Phonon spectra of the Cr along the two main Brillouin Zone directions $\Gamma$H and $\Gamma$N at 0.1~GPa (left) and at 15.6~GPa (right). The open symbols are published neutron data from Ref.~\onlinecite{Shaw1971}, closed symbols are from this work. The solid line are the fit based on the Born-von K\`arm\`an calculation. The dotted line are the ab-initio calculation. The inset of the left panel is the representation of the first Brillouin Zone of a BCC crystal. The black circles are the Bragg peaks, and open are the forbidden reflections.  } \label{Fig2}
		\end{center}
	\end{figure*}	
\section{Experiment}
	
	A single crystal of Cr was cut and chemically etched down to a typical thickness of several dozen micrometers. The mosaic of the sample is 0.05$^\circ$ on the $\boldsymbol{Q}$=(1, 1, 0) and 0.05$^\circ$ on the $\boldsymbol{Q}$=(2, 0, 0) Bragg peaks. The inelastic X-Ray scattering experiment was performed at the beamline ID28 of ESRF in a backscattering configuration. The incident beam was monochromatized thanks to the (9,9,9) reflection of the Si at a wavelength of 0.6968 \AA ~ and the near backscattering configuration warrants an energy resolving power of $\approx$1.5 10$^{-7}$ and an ensuing lorentzian-like energy resolution of  2.6~meV full-width-half-maximum (FWHM). Energy scans were performed with a step of 0.7~meV and acquisition times of 2~min/point. Up to nine different momentum transfers can be recorded simultaneously \cite{Krisch2007}.
	
	The sample was placed in a diamond anvil cell with Helium as the transmitting pressure medium, ensuring excellent hydrostatic conditions. All the measurements reported here have been carried out at room temperature, in the paramagnetic phase. The pressure was measured by the ruby fluorescence method. The drift of the pressure was less than $2\%$ during the entire measurements of the phonon branches (40h). The axis of the pressure cell was placed approximately along the X-ray beam direction. 
 The crystal orientation was chosen such that high symmetry directions of the Brillouin zone (BZ) can be accessed. Phonons were fitted with a damped harmonic oscillator (DHO) convoluted with a lorentzian energy resolution fixed to 2.6~meV. The damping of the DHO is a free parameter but is always far below the FWHM of the lorentzian, typically below 0.3~meV. The wave vectors $\boldsymbol{Q}$ are expressed in relative lattice units of $\|\boldsymbol{a^*}\|$. The lattice parameters were extracted at each pressure from the position of the $\boldsymbol{Q}$=(1, 1, 0) and the $\boldsymbol{Q}$=(2, 0, 0) Bragg peaks. Even at the highest pressure, the mosaïcity of these two Bragg peaks are below 0.06$^\circ$.
 
	\begin{figure}[htbp]
		\begin{center}
			\includegraphics[width = 0.7\columnwidth]{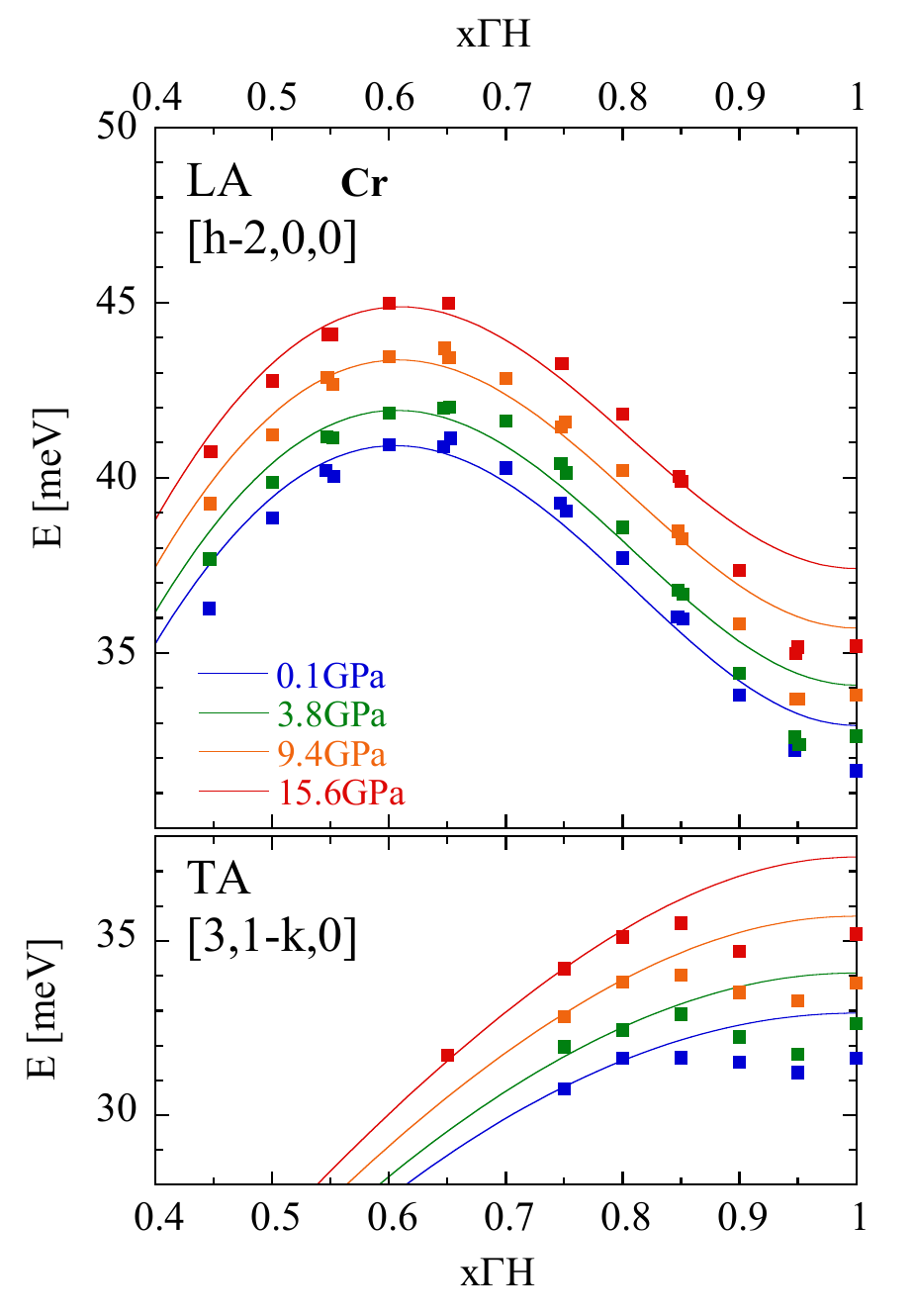}
			\caption{Pressure dependence of the phonon dispersion of the LA (top) and TA (bottom) modes along the $\Gamma H-$direction. The lines are the calculated dispersion curves with the force constants described in the paper.} \label{Fig3}
		\end{center}
	\end{figure}
	Inelastic X-ray measurements consisted in energy scans at constant wave vector, $\boldsymbol{Q}$. Measurements along the $\Gamma H-$direction have been performed at proximity of the $H-$point at  $\boldsymbol{Q}$=(3, 0, 0). From $\boldsymbol{Q}$=(3, 0, 0) to $\boldsymbol{Q}$=(2, 0, 0) for the longitudinal acoustic (LA) mode, and from $\boldsymbol{Q}$=(3, 0, 0) to $\boldsymbol{Q}$=(3, 1 ,0) for the transverse $\parallel \boldsymbol{a^*}$ mode. In Fig. \ref{Fig1} we show typical energy scans of the LA mode propagating along the [1,0,0] direction at position $\boldsymbol{q}$=(0.9, 0, 0), very close to the Brillouin zone boundary, and at different pressures. As expected, a hardening of this mode is clearly observed. The central line is also measured to calibrate the zero-energy position. For the $\Gamma N$ direction, the T2 mode is measured close to the $\boldsymbol{Q}$=(3, 0.5, 0.5), the $N$-point along the direction (3, 1-$l$, $l$). At the maximum pressure, the T1 and L modes have been also measured in the BZ close to the $\boldsymbol{Q}$=(2, 2, 0) Bragg peak along (2+$h$, 2-$h$ ,0) and (2-$h$, 2-$h$, 0) directions, respectively. 
	
	Density Functionnal Theory (DFT) calculations have been performed using \textsc{quantum espresso \cite{Giannozzi2009,Giannozzi2017}}. We use the Perdew–Burke–Ernzerhof (PBE) functional  with pseudopotential from the pseudodojo database. We use a k-point grid of $16\times 16\times 16$. For the calculation of electronic nesting and the Fermi surface, this grid was further interpolated using the Boltztrap2 package \cite{Madsen2018}. The value of the gaussian smearing has a large impact on the phonon frequencies, specially on the lower energy mode (T2) close to the N point (see SI). Moreover, for a large smearing value the Kohn anomalies disappear.  For the calculations shown in Fig. \ref{Fig2}, we are using a smearing of $\gamma=0.05~Ry$. Phonon calculations were performed using Density Functional Perturbation Theory (DFPT) on a q-grid of $16\times 16\times 16$. For the calculation of the electron-phonon coupling, an interpolation of the Brillouin zone was performed.

\section{Results and Discussion} 
	
Complete phonon dispersion curves along $\Gamma H-$  and $\Gamma N-$directions at ambient pressure and at 15.6~GPa, i.e., above the critical pressure, have been measured and are displayed in Fig.~\ref{Fig2}. The dispersion curves are fitted with a Born-von K\`arm\`an model taking into account interactions up to the fifth nearest neighbors. The data from Shaw and Muhlestein\cite{Shaw1971} at ambient pressure have also been included in the fit. 
The force constants extracted from the fit at ambient pressure closely agree with the previously published values \cite{Shaw1971,Trampenau93,Bernal-Choban2023}. 
The \textit{ab initio} calculation of the dispersion curves is also included. Interestingly, the different branches are qualitatively and quantitatively well described. Close to the N point, the T1 branch along the $\Gamma$N direction is slightly overestimated. It is this branch which is strongly dependent on the smearing parameter. Experimentally, this is the same phonon mode that exhibits a strong temperature dependence and is significantly affected by the diffusion jump into a nearest-neighbor vacancy \cite{Trampenau93}. This sensitivity is reproduced by the \textit{ab initio} calculations through the electronic smearing parameter, which affects this branch most strongly. The T2 transverse modes near the $N$ and $H$ points are also particularly sensitive to electronic smearing because their frequencies depend on fine details of the Fermi surface (See SM fig S2).\\

As expected, all acoustic branches harden under pressure and the dispersion curves at 15.6~GPa have been fitted with the same model. The fit was initiated with the force constants obtained at ambient pressure incremented by 20\%. The results are listed in SI. For the intermediate pressures, the tensor force constants are linearly interpolated and the dispersion relation calculated.  

	\begin{table}[htbp]
	    \centering
	    \begin{tabular}{c|cccc}
	        P & a & $c_{11}$    & $c_{44}$ & ($c_{11}-c_{12}$)/2  \\
        (GPa) &  (\AA) & (GPa)  & (GPa) & (GPa)  \\
        \hline
	        0.1 &  2.8885 & 384 & 127 & 131 \\
          3.8  & 2.870 & \textit{409} & \textit{134} & \textit{141} \\
           9.4  & 2.846 & \textit{447} & \textit{144} & \textit{157}  \\
            15.6 & 2.825  & 489 & 155 & 176 \\
            
	    \end{tabular}
	    \caption{Pressure dependence of the lattice parameter and the elastic constants. In italics, the values are calculated from the estimated force constants tensors (see text).   }
	    \label{tabElasticity}
	\end{table}

The values of the propagation velocity of the acoustic modes, and consequently of the elastic constants, are calculated from the dispersion curves. The relationship between a cubic crystal's elastic constants and sound velocities is very well known. Along the $\Gamma H-$direction, $c_{11}=\rho v^2_{L}$ and $c_{44}=\rho v^2_{T}$ whereas along the $\Gamma N$ direction, $c_{44}=\rho v^2_{T2}$ and  $c'=(c_{11}-c_{12})/2=\rho v^2_{T1}$. As usual, $\rho$ is the density and $v_{i}$ the propagation velocity of the mode $i$. The values are shown on the table \ref{tabElasticity}. Elastic constants at ambient pressure and room temperature, measured by ultrasonic techniques by different groups\cite{Bolef1963,Muir1987}, are $c_{11}$=350~GPa, $c_{44}$=100.8~GPa and $c'$=141.1~GPa. $c'$ is comparable to our results, whereas $c_{11}$ and $c_{44}$ are stronger on the phonon dispersion. From ambient pressure to 15.6~GPa, the elastic constants increase between 22 and 34\%. The anisotropy coefficient defined by the ratio $c_{44}/c'$ is anomalously low in Cr compared to the group-IV bcc metals. At ambient temperature and pressure this ratio is below 1, the value which signs an isotropic compound at room temperature. Here, the effect of pressure effect is to harden the lattice and decreases the ratio $c_{44}/c'$. This effect is similar to that of the temperature  where this ratio is increasing when the lattice is softening \cite{Trampenau93}.

The pressure dependence of the longitudinal (LA) and transverse acoustic (TA) modes close to the $H-$point, i.e., close to the positions where the SDW and CDW superstructures appear, are shown Fig.~\ref{Fig3}. The phonon dispersion is calculated with a force tensor linearly interpolated between the ambient pressure and 15.6~GPa. As previously reported, a clear anomalous softening relative to the Born–von Kármán dispersion is visible near the $H$ point. This softening occurs in the vicinity of the nesting vector and is therefore identified as a Kohn anomaly\cite{Lamago10}. The anomaly remains visible at all measured pressures, both below and above the critical pressure. Its relative magnitude and $\boldsymbol{q}$-space extent remain qualitatively unchanged.

\begin{figure}[htbp]
		\begin{center}
            \includegraphics[width = \columnwidth]{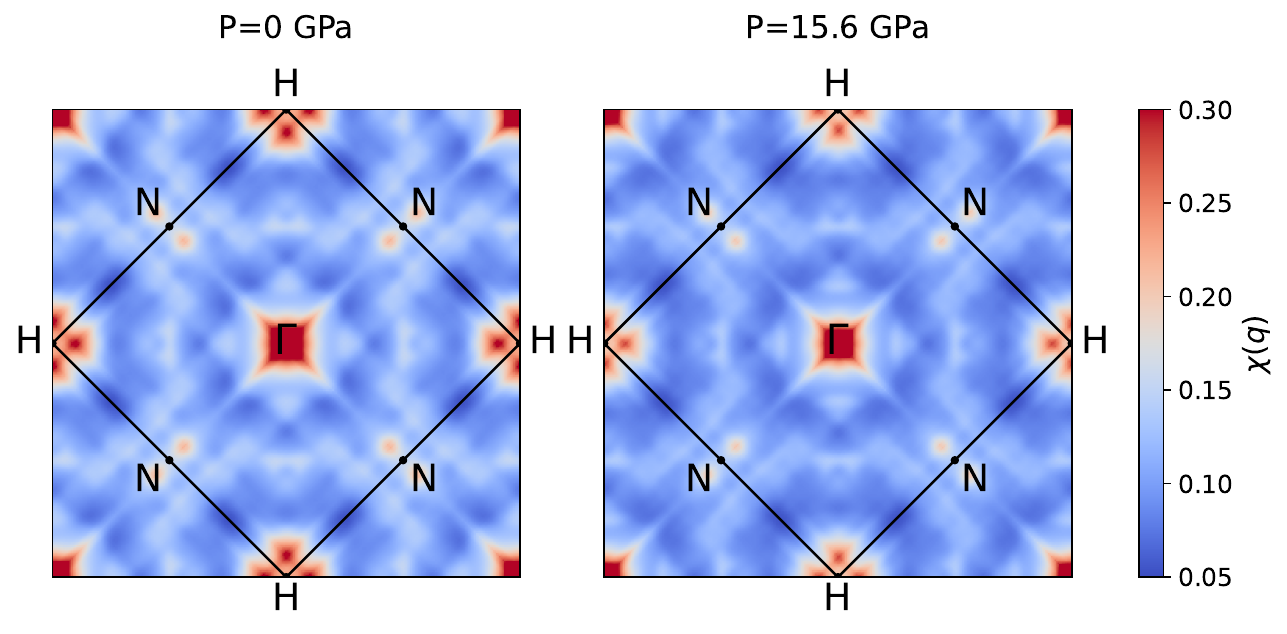}

            \caption{Nesting function in the (hk0) plane at ambient pressure (left) and at 15.6~GPa (right).} \label{Fig4}
		\end{center}
	\end{figure}

To understand the robustness of the Kohn anomalies under pressure, we calculate the Fermi-surface nesting function from first principles. This function is related to the zero frequency limit of the imaginary part of the electronic susceptibility   : 
\begin{equation}
\lim\limits_{\omega\to 0}\chi ''(\mathbf q,\omega)/\omega =  \sum_{m,n} \sum_{\mathbf{k}} \delta(\epsilon_{\mathbf{k+q},m} - \epsilon_F) \delta(\epsilon_{\mathbf{k},n} - \epsilon_F)  
\end{equation}
where $m$ and $n$ denote the different bands, $\delta$ is the Dirac delta function, and $\epsilon_F$ is the Fermi energy. We present the pressure evolution of this function in the $\Gamma H N$ plane in  Fig.~\ref{Fig4}. Remarkably, the maximum remains robust, and no change in the $\mathbf{q}$-vector is observed. These maximum are at the origin of the Kohn anomaly, and thus the robustness of the nesting vectors provides a natural explanation for the Kohn anomalies. Furthermore, our results are in good agreement with recent DFT simulations, which calculated the momentum dependence of the particle-hole susceptibility along specific high-symmetry directions at 10~GPa \cite{Belozerov21}.

Our experimental result is in sharp contrast to other CDW systems, for example 2H-NbSe$_2$, where the energy shift is strongly pressure and temperature dependent at the proximity of the lattice instability \cite{Leroux2015}.

\begin{figure}[tbp]
		\begin{center}
            \includegraphics[width = 1.0\columnwidth]{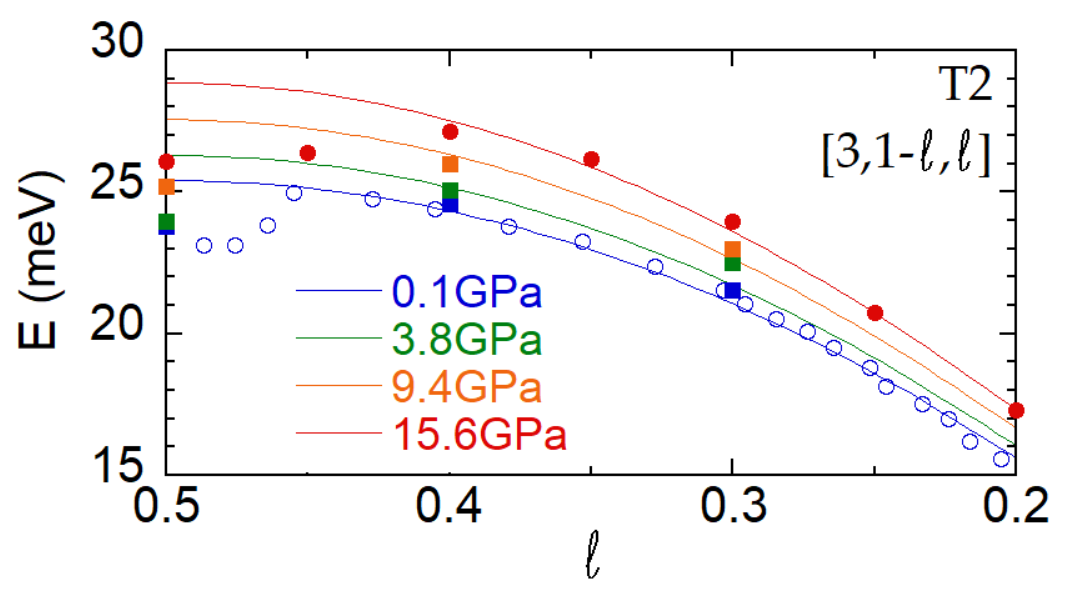}
			\caption{Pressure dependence of the phonon dispersion of the T2 acoustic mode along the $\Gamma$N direction. $\boldsymbol{Q}$=(3, 0.5, 0.5) is a N point. The blue open symbol are from Ref.~\onlinecite{Shaw1971}. The lines are the calculated dispersion curves with the force constants described in the paper.} \label{Fig5}
		\end{center}
	\end{figure}

The pressure dependence of the T2 branch near $N$ is shown in Fig.~\ref{Fig5}. As at $H$, a pronounced anomaly persists over the entire pressure range. The downward deviation from the smooth Born–von Kármán dispersion signals a softening beyond conventional harmonic lattice dynamics. This softening has been attributed to enhanced electron-phonon coupling arising from interband transitions along the $HN$ line \cite{Lamago10}. Although no superstructure is observed near $N$, the anomaly coincides with a local maximum of the nesting function. Its wave vector remains unchanged between 0.1 and 15.6~GPa, while its amplitude decreases slightly.

Our first-principles calculations further connect the electronic structure to the phonon anomalies. The branches displaying the Kohn anomalies near $H$ and $N$ also exhibit strong electron-phonon coupling near the corresponding wave vectors (SM Fig. S1). Upon increasing the electronic smearing, which washes out fine structure near the Fermi level, both the nesting maxima and the Kohn anomalies vanish (SM Figs. S2 and S3).

Under pressure, however, the maximum deviation from the Born–von Kármán dispersion increases slightly, qualitatively consistent with the narrowing of the CDW peak near $P_c$ reported in Ref.\cite{Jaramillo09}. By contrast, the calculated nesting maxima decrease slightly over the same pressure range (Fig. 4). These contrasting trends indicate that the strength of the phonon anomalies reflects the combined influence of Fermi-surface geometry and electron-phonon coupling. Neither anomaly exhibits any critical evolution across $P_c$. This is our main result.

Now, we examine the consequences of the robustness of the Fermi surface nesting on the physics of the CDW/SDW in Chromium and the associated quantum phase transition. It is widely accepted that the critical pressure $P_c$, at which the superstructure peak associated with the CDW reaches 0~K, corresponds to the pressure at which the SDW phase of Chromium is destroyed. The connection between the superstructure peaks and antiferromagnetism is established through the ambient pressure intensity and the $\mathbf{Q}$-vector relationship between the magnetic Bragg peaks measured by neutron scattering and the charge Bragg peaks observed by X-ray diffraction. The mechanism underlying this link remains debated, with proposals including a magnetoelastic mechanism \cite{Tsunoda1974} or a purely electronic coupling \cite{Young74}. The latter scenario could explain the uncorrelated fluctuations in the charge and spin channels \cite{Jacques2014}. Nevertheless, both mechanisms imply that the CDW is induced by the SDW, and thus the SDW ground state exists only below $P_c$ \cite{Jaramillo09}. The robustness of the Fermi surface nesting and the electron-phonon coupling above $P_c$ evidences that another mechanism, such as, for example, a reduction in the exchange coupling, is responsible for this suppression. \\

Beyond chromium as a model system, numerous strongly correlated electron systems---such as cuprates \cite{Tranquada1995, Miao2017,Lee2022} and nickelates \cite{Tranquada1994,Zhang2020}---exhibit both CDW and SDW phases. In these systems, the $\mathbf{Q}$ vector of the CDW can sometimes be twice that of the SDW, though this relationship is not systematic. This variability may arise from a high sensitivity to extrinsic effects, such as CDW pinning. Additionally, phonon anomalies have an unusual behavior, possibly due to an hybridization with CDW excitations \cite{Miao2018, Souliou2025}. The relationship between the Fermi surface, Kohn anomalies, and coupled density waves in weakly correlated chromium could thus serve as a reference for understanding these more complex systems.

\section{Conclusion}

Our results for the phonon dispersion curves under pressure at room temperature, combined with \textit{ab initio} calculations, support the idea that a robust Fermi surface nesting and electron-phonon coupling renormalize the phonon dispersion at the $H$- and $N$-points. However, they do not play a direct role in the charge density wave ordering in chromium or in the critical pressure. At least not in the way that is expected from a Peierls-like mechanism where a phonon mode softens at the approach of the phase transition and this condensation leads to a new superstructure at 2$k_F$. Different types of (DFT-DMFT) calculations \cite{Schafer17,Belozerov21} have underlined the need of the so-called Kohn points to yield a maximum of the non-uniform magnetic susceptibility at the SDW modulation wavevector. Our data hereby support the claim that the formation of a SDW is the actual order parameter in the antiferromagnetic phase of Cr, the formation of a CDW is collateral to the appearance of a SDW. The ensuing QCP that develops under pressure or V doping is hardly affected by the Kohn anomalies which are robust upon pressure or temperature.

\section{Acknowledgment}
 
	We acknowledge fruitful discussions with  P. Monceau, M. De Boissieu, R. Heid and V. Jacques. P.R. thanks Manolo Nunez-Regueiro and Sitaram Ramakrishnan for enthusiastic discussions. We thank the ESRF to provide us some beamtime under the number HC-4046. We acknowledge the support of the high pressure service at the ESRF. This work is supported by the ANR-DFG Grant No. ANR-18-CE92-0014-03 “Aperiodic.”

\bibliography{Cr_pressure_V3}
 
\end{document}